\documentclass[acmtog, authorversion]{acmart}
\acmSubmissionID{1234}

\setcitestyle{nosort,square}

\setcopyright{none}
\renewcommand\footnotetextcopyrightpermission[1]{}
\usepackage{bm}
\usepackage{amsfonts}
\usepackage{enumitem}
\usepackage[utf8]{inputenc}
\usepackage{setspace}
\usepackage{listings}
\usepackage{wrapfig}
\usepackage{nicefrac}

\usepackage{caption}
\usepackage{subcaption}
\usepackage{hyperref}

\usepackage{tcolorbox}
\usepackage[newfloat, frozencache]{minted}
\usepackage{tcolorbox}
\usepackage{minted}
\definecolor{hlcol}{RGB}{255,200,200}

\newcommand{\mintedpseudocode}[1]{
    \begin{tcolorbox}[arc=1pt,colback=black!3!white,colframe=black!16!white,boxrule=0.3mm,before skip=1mm,after skip=1mm,top=.5mm,bottom=.5mm,left=1mm,right=1mm]
         \inputminted[escapeinside=||,mathescape=true,fontsize=\footnotesize,linenos]{python}{#1}
    \end{tcolorbox}
}

\newenvironment{code}{\captionsetup{type=listing}}{}
\SetupFloatingEnvironment{listing}{name=Algorithm}

\newcommand{\keepvalues}{%
  \edef\restorevalues{%
    \parindent=\the\parindent
    \parskip=\the\parskip
  }%
}

\definecolor{blue}{rgb}{0.180, 0.180, 0.902}
\definecolor{darkblue}{rgb}{0.17, 0.24, 0.31}

\newcommand{\param}{\pi}
\newcommand{\paramv}{\bm{\param}}

\newcommand{\pathsp}{\mathcal{X}}
\newcommand{\walkpath}{x}

\newcommand{\ball}{B(x)}

\newcommand{\surface}{\partial B(x)}
\newcommand{\surfacek}[1][k]{\partial B(\vx_#1)}
\newcommand{\boundary}{\partial \Omega}
\newcommand{\eshell}{\Omega_{\varepsilon}}

\newcommand{\majorant}{\bar{\sigma}}

\newcommand{\greenS}{G^{\sigma}(x,y)}

\newcommand{\poisson}{P^{\sigma}(x,z)}

\newcommand{\pgrad}{\partial_{\param}}

\newcommand{\unitsphere}{\mathcal{S}^2}

\newcommand*\diff{\mathop{}\!\mathrm{d}}

\DeclareMathOperator*{\argmin}{arg\,min}

\newcommand{\bomega}{\bm{\omega}}

\newcommand{\vx}{\mathbf{x}}
\newcommand{\vy}{\mathbf{y}}

\acmJournal{TOG}

\begin{document}
\title{Solving Inverse PDE Problems using Grid-Free Monte Carlo Estimators}

\author{Ekrem Fatih Yilmazer}
\email{ekrem.yilmazer@epfl.ch}
\affiliation{%
  \institution{École Polytechnique Fédérale de Lausanne (EPFL)}
  \city{Lausanne}
  \country{Switzerland}
}

\author{Delio Vicini}
\email{delio.vicini@epfl.ch}
\affiliation{%
  \institution{École Polytechnique Fédérale de Lausanne (EPFL)}
  \city{Lausanne}
  \country{Switzerland}
}
\author{Wenzel Jakob}
\email{wenzel.jakob@epfl.ch}
\affiliation{%
  \institution{École Polytechnique Fédérale de Lausanne (EPFL)}
  \city{Lausanne}
  \country{Switzerland}
}

\begin{abstract}
    Modeling physical phenomena like heat transport and diffusion is
    crucially dependent on the numerical solution of partial differential
    equations~(PDEs). A PDE solver finds the solution given
    coefficients and a boundary condition, whereas an \emph{inverse PDE solver}
    goes the opposite way and reconstructs these inputs from an existing
    solution. In this article, we investigate techniques for solving inverse
    PDE problems using a gradient-based methodology.

    Conventional PDE solvers based on the finite element method require a
    domain meshing step that can be fragile and costly. Grid-free Monte Carlo
    methods instead stochastically sample paths using variations of the
    \emph{walk on spheres} algorithm to construct an unbiased estimator of the
    solution. The uncanny similarity of these methods to physically-based
    rendering algorithms has been observed by several recent works.

    In the area of rendering, recent progress has led to the development of
    efficient unbiased derivative estimators. They solve an adjoint form of the
    problem and exploit arithmetic invertibility to compute gradients using a
    constant amount of memory and linear time complexity.

    Could these two lines of work be combined to compute cheap parametric
    derivatives of a grid-free PDE solver? We investigate this question and
    present preliminary results.
\end{abstract}

\ccsdesc[500]{Mathematics of computing~Partial differential equations}
\ccsdesc[500]{Computing methodologies~Rendering}

\keywords{walk on spheres, Monte Carlo, differentiable simulation, path replay backpropagation}

\maketitle

\section{Introduction}
Many physical phenomena are naturally described using partial differential
equations~(PDEs). For example, the heat equation models the spread of thermal
energy in a potentially heterogeneous material. Solvers that numerically
approximate solutions of such PDEs are in widespread use. We pursue the
opposite direction in this article, which is known as an \emph{inverse PDE
problem}: estimating unknown parameters from observations of the solution.
This set of unknown parameters could include various PDE coefficients,
boundary conditions, and even the shape of the domain.

Such problems arise in diverse scientific and engineering contexts, for example
to determine the physical parameters of a thermal conductor from
measurements~\cite{Cannon1964}. Electrical impedance
tomography~\cite{Cheney1999} seeks to reconstruct the interior of a living
organism. Electrodes provide measurements of the electric field, which is
influenced by the tissue's conductivity, impedance, and dielectric permittivity.

Our approach entails differentiating the solver and recovering the unknown
parameters using gradient descent. However, one issue with conventional PDE
solvers based on the finite element method~(FEM) is that they require a meshing
step that can be fragile and computationally costly. An alternative are Monte
Carlo PDE solvers based on the \emph{walk on spheres} (WoS)~\cite{Muller1956}.
These \emph{grid-free} methods sample random paths in the domain to compute
unbiased estimates of the solution. Grid-free solvers have recently attracted
significant attention in the computer graphics community, partly owing to the
remarkable similarities to Monte Carlo rendering methods~\cite{Rohan20} and the
algorithmic synergies that this creates~\cite{Rohan22,Qi22bidirectional}.

A common issue with gradient-based optimization is that the standard approach
for reverse-mode differentiation (known as \emph{backpropagation}) reverses all
data dependencies of an underlying computation. When applied to the WoS
algorithm, this means that intermediate results of a large number of iterations
would need to be stored to enable the subsequent differentiation.

In the field of rendering, recent progress has led to the development of
\emph{differentiable rendering}
methods~\cite{Gkioulekas2013,Li18EdgeSampling,NimierDavidVicini2019Mitsuba2}
that estimate parametric derivatives of complete light transport simulations. A
similar issue arises here as well: light paths can potentially be very long,
particularly in highly-scattering media, which makes na\"ive reverse-mode
differentiation prohibitively memory-intensive. Methods like \emph{radiative
backpropagation}~\cite{NimierDavid2020Radiative} and \emph{path replay
backpropagation}~\cite{Vicini2021PathReplay} cast the differentiation step into
an independent simulation of ``derivative light'' to address this issue. The
latter project solves an adjoint version of the underlying equation and
furthermore exploits arithmetic invertibility in the computation to
differentiate using a constant amount of memory and a runtime cost that is
linear in the number of path vertices.

Given these striking similarities, could a similar approach be useful to compute
reverse-mode derivatives of grid-free Monte Carlo solvers? We show that this is
indeed the case and that this combination yields an unbiased derivative
estimator in the same complexity class. The paper presents preliminary results
on synthetic inverse problems. We make no claims about the utility of such an
approach for solving concrete inverse-PDE problems but find it a promising
direction for future work.

\section{Method}
\label{section:estimators}
\subsection{Background}
\paragraph{Inverse PDE problems.}
We seek to solve an inverse PDE problem of the form:
\begin{align}
    \widehat{\paramv} = \argmin_{\paramv} \ell (u(\paramv)),
\end{align}
where $u(\paramv)$ is the solution of a PDE parameterized by the vector
$\paramv$ containing the boundary values, source terms, etc. The function $\ell$
is a differentiable objective function. In the simplest case, this could be the
$L_2$ difference between the solution of the PDE and a reference solution
evaluated at a set of locations spread throughout the domain. We use a regular
grid in experiments, though irregularly spaced evaluation points are also within
the scope of this approach.

To attempt to solve this problem using gradient descent, we must differentiate
the objective $\ell$ (and therefore, also the solver $u(\paramv)$) with respect
to $\paramv$. For notational simplicity, we shall focus on a derivative
$\partial_{\param} \coloneq \partial / \partial\param$ with respect to a single
parameter $\param$, though the resulting methods also generalize to arbitrary
parameter counts.

\paragraph{Primal solver.} The walk on spheres method~\cite{Muller1956}
expresses the solution $u(x, \param)$ as a recursive integration problem. In
this paper, we consider various PDEs which have a solution expressed as a
\emph{Fredholm integral equation of the second kind}:
\begin{align}
    u(x, \param) = S(x, \param) + \int_{\mathcal{Y}} K(x, x', \param) \,u(x, \param) \diff x'.
    \label{eq:fredholm}
\end{align}
Here, $\mathcal{Y}$ is typically a sphere around $x$ and $K$ attenuates the
recursive contribution of $u$. The term $S$ relates to the \emph{source term}
and is usually itself an integral that, however, does not reference the solution
$u$. We assume both $S$ and $K$ to potentially depend on a differentiable
parameter~$\param$. This formulation covers the range of PDEs described by
recent works in computer graphics~\cite{Rohan20, Rohan22}. The solution of the
PDE can then efficiently be estimated using recursive Monte Carlo integration.
We refer to~\citeauthor{Rohan22}~\shortcite{Rohan20, Rohan22} for an in-depth
introduction and discussion of walk on spheres.

This recursive formulation is reminiscent of the rendering equation~\cite{kajiya1986}
solved by physically-based rendering algorithms:
\begin{equation}
    L_o(\vx, \bomega) = L_e(\vx, \bomega) + \int_{\unitsphere} f_s (\vx,\bomega, \bomega') L_i(\vx, \bomega') \diff \bomega'^{\bot},
    \label{eqn:rendering}
\end{equation}
where $L_o$, $L_i$ and $L_e$ denote the outgoing, incident and emitted radiance
and $f$ is the \textit{bidirectional scattering distribution function} (BSDF).
In both equations, the solution appears recursively in the integral on the
right. Both walk on spheres and rendering algorithms need to recursively sample
the integrand on the right, potentially constructing long paths before
terminating at a boundary or on a light source.

\paragraph{Derivative integrals.}
Instead of using a standard framework for automatic differentiation that would
record the computational structure of the random walk at significant expense,
recent differentiable rendering methods~\cite{NimierDavid2020Radiative,
Vicini2021PathReplay} cast the differentiation step into an independent random
walk that propagates differential quantities (e.g., the derivative of radiance)
through the domain.

These methods can be motivated by formulating the derivative of the solution by
applying the derivative operator to Equation~\ref{eq:fredholm}. Under the
assumption that the parameter $\param$ does not affect the position of
discontinuities in the integrand or the integration domain $\mathcal{Y}$ itself,
we then get:
\begin{align}
    \pgrad u(x, \param) &= \pgrad S(x, \param) + \pgrad \int_{\mathcal{Y}} K(x, x', \param) \,u(x, \param) \diff x' \nonumber\\
                        &= \pgrad S(x, \param) + \int_{\mathcal{Y}} \pgrad \left[K(x, x', \param)\right] u(x, \param) \diff x' \nonumber\\
                        & \hspace{5em} + \int_{\mathcal{Y}} K(x, x', \param) \, \pgrad u(x, \param) \diff x'.
    \label{eq:diff_fredholm}
\end{align}

In contrast to the primal problem, we now have \emph{two} recursive terms to estimate:
both $\pgrad u$ and $u$ require a separate random walk. In differentiable
rendering,
\citeauthor{NimierDavid2020Radiative}~\shortcite{NimierDavid2020Radiative}
suggested to recursively estimate both quantities, which results in quadratic
complexity. We will instead use the path replay backpropagation
algorithm~\cite{Vicini2021PathReplay}, which obtains the derivative in linear
time.

In this paper, we will assume that the domain of the PDEs we solve is fixed. If
that were not the case, $\mathcal{Y}$ might be parameter dependent. In the
simplest case, $\mathcal{Y}$ is the largest sphere around $x$, which changes as
the domain itself changes. We could then not simply move the derivative operator
into the integral. We also assume that the functions $S$, $K$, as well as
Dirichlet boundary values, are continuous. The discontinuous case has been
studied in differentiable rendering and requires specialized techniques such as
\emph{edge sampling} \cite{Li18EdgeSampling},
\emph{reparameterizations}~\cite{Loubet2019Reparameterizing,
bangaru2020warpedsampling}, or path-space edge sampling~\cite{Zhang:2020:PSDR}.
We sidestep this point and focus on integrands that are continuous with respect
to $\param$. We leave the application of such techniques to PDEs with
discontinuous parameters for future work.

\subsection{Path replay backpropagation}
\label{sec:prb}
In this paper, we adapt path replay backpropagation~\cite{Vicini2021PathReplay}
that applies to differential estimators of both Equations~\ref{eq:fredholm} and
\ref{eqn:rendering}. The method is best explained using pseudocode.
Algorithm~\ref{alg:fredholm} implements a simple primal solver for
Equation~\ref{eq:fredholm}:
\begin{code}%
    \mintedpseudocode{code/fredholm_solver.py}
    \caption{
        \label{alg:fredholm}
        Monte Carlo solver for a Fredholm equation of the second kind.
    }
\end{code}%
It computes a one-sample estimate of the solution at \verb|x|. Initially, the
running estimate $u$ is set to 0 and the \emph{throughput} $\beta$ is set to~1.
In each iteration, we accumulate the throughput-weighted source contribution
(e.g., source term of the PDE or surface emission in rendering) and then sample
the next position \verb|x|, while updating the throughput using the ratio of $K$
and the sampling PDF. In rendering, this would be the ratio of the BSDF and the
solid angle density of the sampling strategy. The evaluation of $S$ might also
involve a Monte Carlo estimator, but this is usually simple as it does not
recursively depend on $u$. If we were to use next event estimation, we would
also consider it to be part of the evaluation of $S$ in this context.

Path replay backpropagation can now be used to estimate derivatives in reverse
mode. The central idea of this approach is to estimate the solution $u$ and
subsequently accumulate gradients by regenerating the \emph{same path} a second
time. In a concrete implementation, this would typically be accomplished by
re-seeding the underlying source of pseudorandomness. When the path is
reconstructed once more, the method can leverage the known final result to
compute an unbiased gradient estimate. This involves no approximations and
removes the need to store large temporary buffers of intermediate random walk
state.

The method evaluates a vector-Jacobian product $\delta_{\param} =
\delta_u^T J_{s}$, where $J_{s}$ is the Jacobian of the solver in
Algorithm~\ref{alg:fredholm}. The vector~$\delta_u$ contains the gradient of the
objective function with respect to the value of $u$, and $\delta_{\param}$ is the
desired vector of parameter derivatives for gradient descent.

Algorithm~\ref{alg:fredholm_grad} implements path replay backpropagation for
the general solver in Algorithm~\ref{alg:fredholm}:
\begin{code}%
    \mintedpseudocode{code/fredholm_solver_grad.py}
    \caption{%
        \label{alg:fredholm_grad}%
        Path replay version of the Fredholm equation solver.}
\end{code}%
The function takes the query location $x$, the output of the first pass $u$ and
the objective function derivative $\delta_u$ as input. Here, we both evaluate
the parameter-dependent functions~$S$ and~$K$ using automatic differentiation,
and the function \verb|backward| then backpropagates gradients into their
parameters. By default, automatic differentiation eagerly propagates gradients
through all involved quantities, but here we actually want to localize the usage
of AD to consider only part of the loop body. Any computations outside the
\verb|ad_enabled()| blocks will not be differentiated by automatic
differentiation. This also means that we never build an automatic
differentiation graph across loop iterations, and hence do not need to allocate
memory to store all the loop states. Note that the above pseudocode evaluates
certain terms twice (e.g.,~$S(x, \param)$). This is purely for notational
clarity and in practice the same logic can be implemented using a single
forward evaluation of these terms.

The above algorithm computes
\emph{detached}~\cite{Zeltner2021MonteCarlo,Vicini2021PathReplay} gradients,
which means that we estimate the gradients of a parameter-dependent integral as:
\begin{align}
    \int_{\pathsp} \pgrad f(\walkpath, \param) \diff \walkpath \approx \frac{1}{N} \sum_{i=1}^{N} \frac{\pgrad f(\walkpath_i, \param)}{p(\walkpath_i, \param)},
\end{align}
where $N$ is a number of samples and $\walkpath_i$ are i.i.d. samples with the
probability density function $p(\walkpath_i)$. This estimator is detached in a
sense that we do not differentiate the sampling strategy and PDF. We only
differentiate the integrand $f$ itself. On the other hand, \emph{attached}
estimators differentiate the sampling strategy, which is more expensive and
requires additional precautions due to
discontinuities~\cite{Zeltner2021MonteCarlo}. In rendering, their use is limited
to special cases such as perfectly specular surfaces. We use detached estimators
throughout this paper, but attached estimators could potentially be useful to
handle gradients with respect to the domain boundary.

\subsection{Poisson equation}
\label{sec:poisson}
The remainder of this article discusses different PDEs in turn: the Poisson
equation (this section), its screened variant
(Section~\ref{sec:screened_poisson}), and the general case of a heterogeneous
elliptic PDE (Section~\ref{sec:elliptic}).

The Poisson equation is defined as
\begin{alignat}{3}
    \Delta u(x)  &= -f(x) \quad   &&x\in\Omega, \nonumber \\
    u(x)         &= g(x)  &&x\in\boundary,
    \label{eqn:poisson}
\end{alignat}%
where $\Omega$ is the domain, $\boundary$ its boundary and $u$ the solution
function. The function $f(x)$ is a (spatially-varying) source term and $g(x)$
models the boundary values.

\paragraph{Primal estimator.}
The solution of the Poisson equation satisfies the following recursive
equation~\cite{Rohan20, Delaurentis1990}:
\begin{equation}
    u(x) = \underbrace{\int_{B(x)} f(y) G(x,y) \diff y}_{\eqqcolon S} + \underbrace{\frac{1}{|\surface|}}_{\eqqcolon K} \int_{\surface} u(z) \diff z,
\end{equation}
where $G(x, y)$ is the Green's function. Walk on spheres estimates $u(x)$
recursively by repeatedly sampling a point on a sphere~$\surface$ around the current
position until it reaches the boundary. For efficiency, we use the largest
sphere that is still entirely contained inside the domain. Additionally, in each
step we sample a position $y$ inside the current sphere to evaluate the source
term~\cite{Rohan20}. This process is illustrated in
Figure~\ref{fig:poisson_primal}. A single sample of the walk on spheres
estimator can be written as:
\begin{figure}
    \includegraphics[width=0.8\columnwidth]{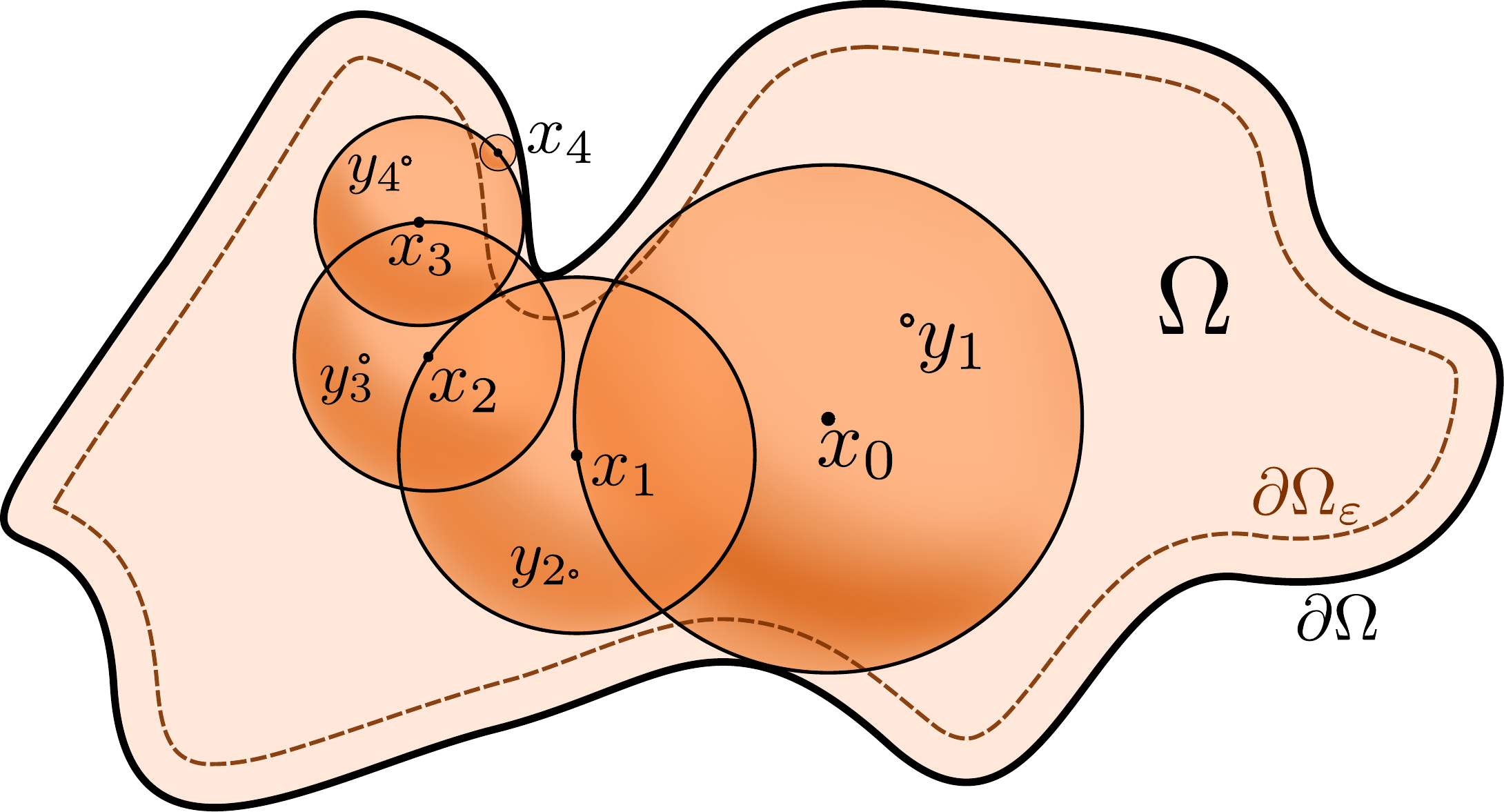}
    \centering
    \caption{Illustration of the steps taken by the walk on spheres estimator
     for the Poisson equation. Starting from a point $x_0$, we repeatedly sample
     the next point $x_i$ on the sphere around the current point until we sample a
     point inside~$\eshell$. The set~$\eshell$ contains the points within
     $\varepsilon$-distance to the boundary $\boundary$. Once we sample such a
     point, we evaluate the Dirichlet boundary values $g(x)$ and terminate the
     random walk. In each iteration, we further sample a point $y_i$ to evaluate the source
     term $f$.}
    \label{fig:poisson_primal}
\end{figure}
\begin{align}
    u(x) \approx \sum_{k=1}^N f(y_k) |G(\walkpath_k)| + g(\walkpath_N),
    \label{eq:poisson_estim}
\end{align}
where $|G(\walkpath_k)| = \int_{B(\walkpath_k)} G(\walkpath_k, y) \diff y$ and
$g(\walkpath_N)$ is the evaluation of the boundary condition at the point where we reach the domain boundary. We sample $y_k \in
B(\walkpath_k)$ proportionally to the Green's function and $\walkpath_{k+1} \in
\surfacek$ uniformly, which caused the terms $G(\walkpath_k,\vy_k)$ and
$1/|\surfacek|$ to cancel in the above expression. Pseudocode for this estimator is shown in
Algorithm~\ref{algpoisson}. The function \verb|poisson| estimates $u(x)$ for a
given~$x$, source term~$f$ and boundary values~$g$. We compute the distance to
the domain boundary inside \verb|distance_to_boundary()|. The function
\verb|sample_green()| samples \verb|y| proportionally to $G(x, y)$ and
\verb|sample_boundary()| uniformly samples the surface of the ball $B(x)$ of
radius \verb|R|.
\begin{code}%
    \mintedpseudocode{code/wos_poisson.py}
    \caption{
        \label{algpoisson}
        Walk on spheres for the Poisson equation.}
\end{code}%

\begin{figure}
    \includegraphics[width=0.8\columnwidth]{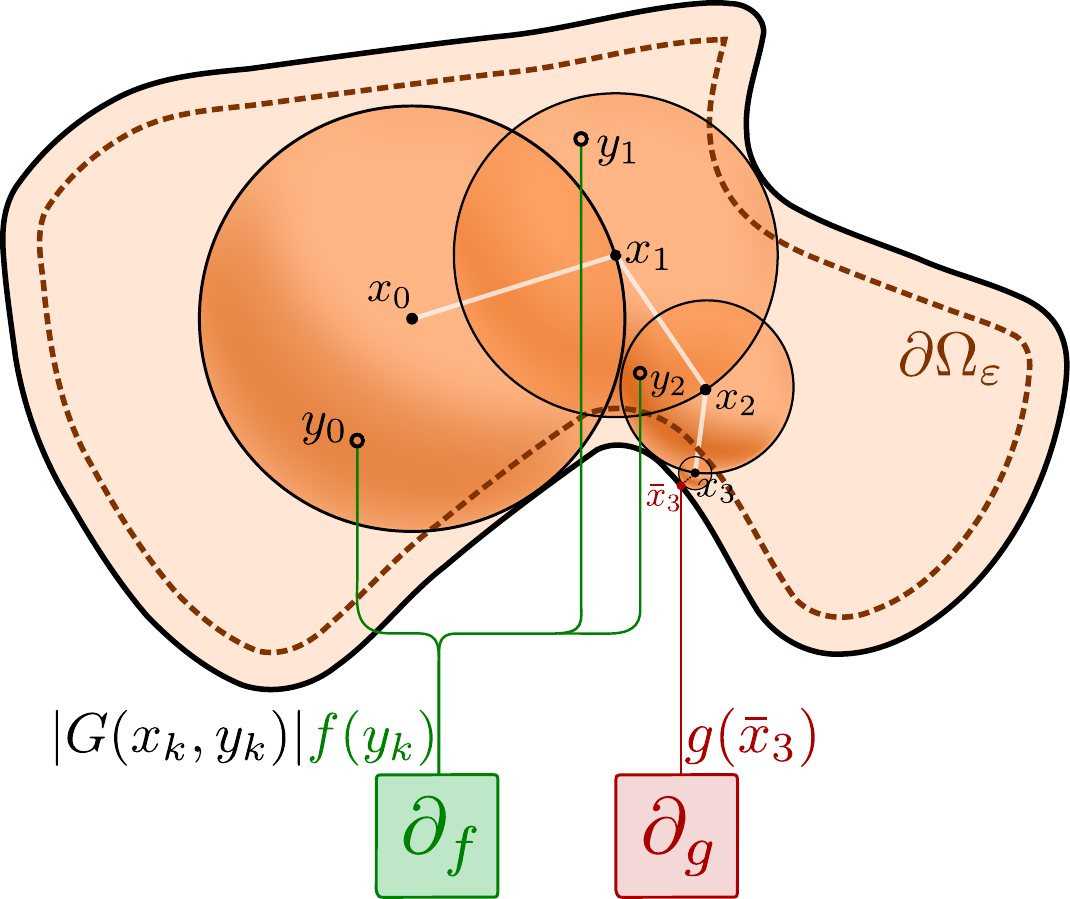}
    \centering
    \caption{Gradient computation for the Poisson equation. The colored (green,
    red) quantities are computed locally using automatic differentiation.}
    \label{fig:poissonGrad}
\end{figure}

\paragraph{Differential WoS}
The Poisson equation solver is the easiest to differentiate. Later sections will
require the use of \emph{path replay} to regenerate a random walk twice, but it
is not needed here. This is because the throughput of the sampled path is not
parameter dependent.

The estimator in Equation~\ref{eq:poisson_estim} computes a weighted sum of $f$
and $g$-evaluations. To differentiate it, we must evaluate the reverse-mode
derivative of all terms in this sum. Algorithm~\ref{algpoisson_grad} implements
this computation using a \emph{differential walk on spheres} (dWoS), which is
illustrated in Figure~\ref{fig:poissonGrad}.

\begin{listing}%
    \mintedpseudocode{code/wos_poisson_grad.py}
    \caption{
        \label{algpoisson_grad}
        Differential walk on spheres for the Poisson equation.}
\end{listing}%

\subsection{Screened Poisson equation}
\label{sec:screened_poisson}
The \emph{screened} Poisson problem presents a slightly more complex case. This
PDE includes an extra
scalar screening coefficient $\sigma$:
\begin{alignat}{3}
    \Delta u(x) - \sigma u(x) &= - f(x) \quad   &&x\in\Omega, \nonumber \\
    u(x)         &= \text{  }g(x)  &&x\in\boundary.
\end{alignat}
Its solution satisfies the following Fredholm integral equation~\cite{Rohan20, Elepov1969}:
\begin{align}
    u(x) =  \underbrace{\int_{\ball} f(y) \greenS \diff y}_{\eqqcolon S} + \int_{\surface} \underbrace{\poisson}_{\eqqcolon K} u(z)\diff z,
    \label{eq:spoisson_solution}
\end{align}
where $\poisson$ is the Poisson kernel and $\greenS$ is the harmonic Green's
function of $B(x)$ (see Appendix~\ref{appendix:green} for the definition of these functions). Both the
Green's function and the Poisson kernel depend on the screening
coefficient~$\sigma$. The sampling strategy remains unchanged: we sample the
Green's function to estimate the source term and draw uniform locations on the
sphere boundary to obtain the next path vertex. However, the value of the
Poisson kernel does not cancel with the sampling PDF and we thus end up with a
parameter-dependent throughput term.

\begin{figure}
    \includegraphics[width=0.75\columnwidth]{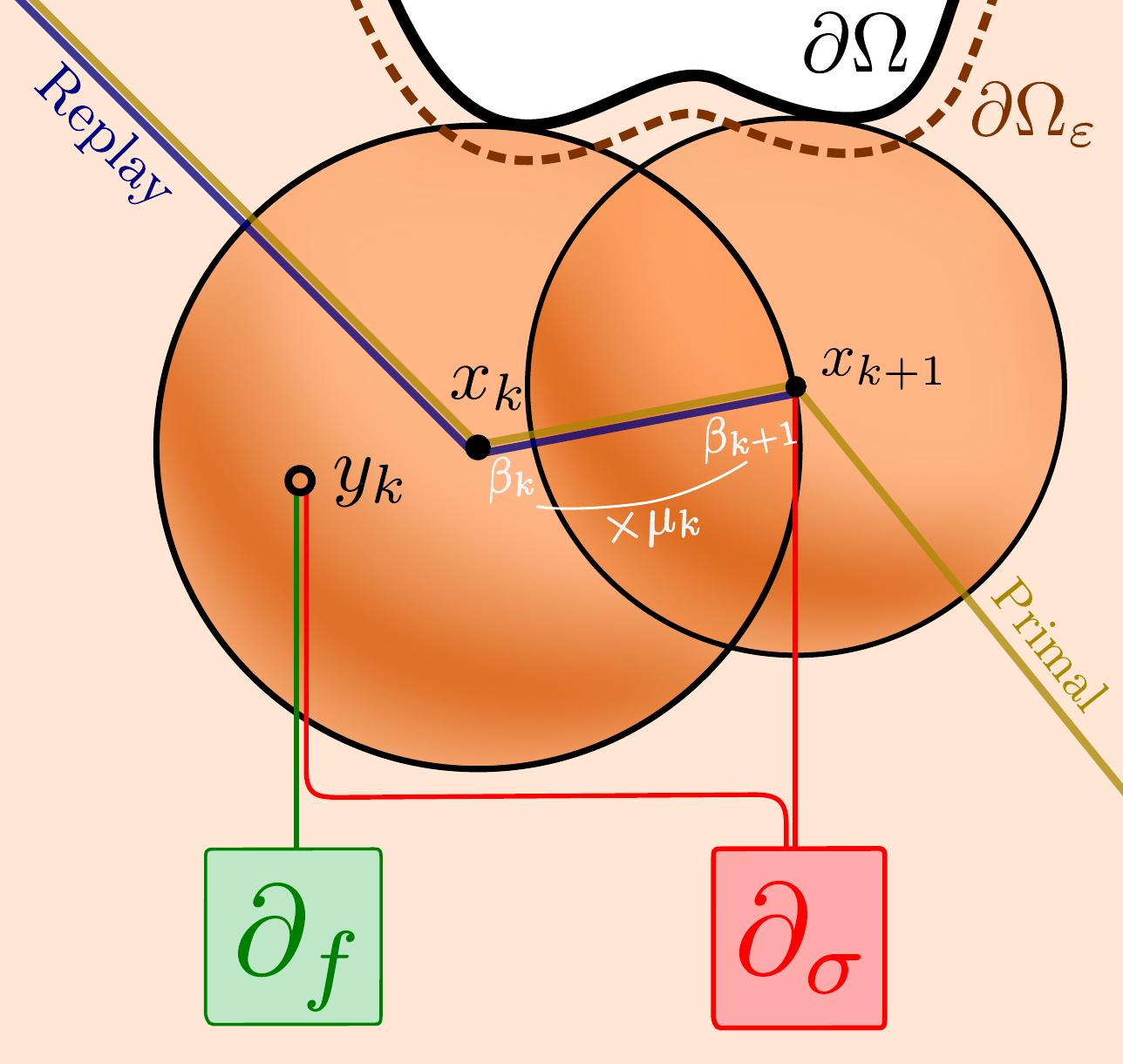}
    \centering
    \caption{Path replay WoS algorithm of the screened Poisson equation (k'th iteration).
    The colored (green, red) lines indicate the positions where the gradients are accumulated.}
    \label{fig:poissonReplay}
\end{figure}

Path replay is now necessary to efficiently differentiate this estimator in
reverse mode. Matching pseudocode is shown in Algorithm~\ref{alg:poissonReplay}.
The structure largely resembles Algorithm~\ref{alg:poissonReplay}, but the
function now takes the result of a previous primal solution estimate as input
(variable \verb|u|). In each step, the current source term evaluation is
subtracted from \verb|u| (Line 15), which is then used to compute the gradient
of the throughput update (Line 20). The quantities used during the reverse-mode
gradient computation are illustrated in Figure~\ref{fig:poissonReplay}.

\begin{listing}%
    \mintedpseudocode{code/wos_sPoisson.py}
    \caption{
        \label{alg:poissonReplay}
        Differential walk on spheres with path replay for the screened Poisson equation.}
\end{listing}%

\subsection{General heterogeneous elliptic problems}
\label{sec:elliptic}
The screened Poisson problem required the use of path replay to estimate
gradients of the screening coefficient. However, the use of path replay for this
problem was mainly pedagogical: since the screening coefficient is just a scalar,
one could easily estimate its derivative using forward-mode AD or even finite
differences.

\citeauthor{Rohan22}~\shortcite{Rohan22} propose WoS-based solvers supporting spatially
varying PDE coefficients building on methods from volumetric transport. We
analyze their delta tracking-based approach and derive the corresponding
differential estimator. The algorithms target elliptic PDEs of the form

\begin{align}
    \nabla (\alpha (x) \nabla u(x)) + \vec{\omega}(x)  \nabla u(x) -\sigma(x)u(x) &= -f(x) && x \in \Omega \nonumber \\
    u(x) &= g(x) && x \in \partial \Omega.
    \label{eqn:elliptic}
\end{align}
Here, $\alpha(x)$, $\omega(x)$ and $\sigma(x)$ are spatially variable diffusion,
transport and screening coefficients. Similar to
\citeauthor{Rohan22}~\shortcite{Rohan22}, we also assume $\vec{\omega}(x) = 0$
for simplicity. That said, the general principles shown here could be used to
develop primal and differential estimators that furthermore account for this
term.

The integral formulation for this general elliptic PDE is:
\begin{align}
    \label{eqn:ellipticSoln}
    u(x) &= \int_{B(x)} f(y) \sqrt{\frac{1}{\alpha(x) \alpha(y)}} G^{\bar{\sigma}}(x,y) \diff y \nonumber &&\text{(T1)}\\
    &+\int_{B(x)} (\bar{\sigma} -  \sigma'(z))  \sqrt{\frac{\alpha(z)}{\alpha(x)}} u(y)  G^{\majorant}(x,z) \diff z \nonumber &&\text{(T2)} \\
    &+\int_{\surface}   \sqrt{\frac{\alpha(z)}{\alpha(x)}} u(z)  P(x,z) \diff y  &&\text{(T3)}
\end{align}
where $\majorant$ is a non-zero fictitious screening coefficient and $\sigma'$ is defined as:
\begin{align}
    \sigma'(x) &\coloneq \frac{\sigma(x)}{\alpha(x)} + \frac{1}{2} \left[
        \frac{\Delta\alpha(x)}{\alpha(x)} - \frac{1}{2}|\nabla \log (\alpha(x))|^2 \right]
\end{align}
This integral form is similar to the integral formulation of delta
tracking~\cite{woodcock1965,galtier2013integral}. When estimating $u(x)$ using
walk on spheres, the source term (T1) is evaluated as before. However, similar
to delta tracking, only one of the terms T2 and T3 is sampled in each step. This
is crucial to avoid quadratic complexity in the number of interactions.
\citeauthor{Rohan22}~\shortcite{Rohan22} derived suitable sampling probabilities
for each term.

\paragraph{Differential WoS}
Since in each step we only sample one of the recursive terms in
Equation~\ref{eqn:ellipticSoln}, we can apply path replay just like before.
During the gradient computation phase, we will sample both terms in the exact
same sequence as during the primal phase.

Intuitively, the discrete sampling decision to either sample T2 or T3 might
appear problematic, as it is non-differentiable. However, our \emph{detached}
estimator does not need to differentiate the sampling process itself. As the
integrand remains continuous, there is no issue differentiating the general
elliptic WoS estimator. In rendering, the same reasoning was used to
differentiate delta tracking using path replay~\cite{Vicini2021PathReplay}.

\begin{listing}%
    \mintedpseudocode{code/wos_delta.py}
    \caption{
        \label{alg:deltaReplay}
        Differential walk on spheres using delta tracking for the general elliptic PDE.}
\end{listing}%

The pseudocode of the algorithm is given in Algorithm~\ref{alg:deltaReplay}.
Analogous to the differential WoS for the screened Poisson equation, the
function uses the previous solution estimate \verb|u|. As before, we subtract
the source term contribution from it in each iteration (Line 15). In Line 18 we
decide whether to sample the volume term T2 or the area term T3. The function
\verb|sampler.rand()| returns a random number that is uniformly distributed in
$[0,1)$. As suggested by \citeauthor{Rohan22}~\shortcite{Rohan22}, we sample the
volume term with a probability $|G(x)|\bar{\sigma}$. The new position
\verb|x_new| is then sampled by either sampling the Green's function or
uniformly sampling of the sphere surface. In the following lines 26-30, we
compute the multiplicative throughput change $\mu$ with AD enabled. In the
notation of the original Fredholm equation solver, $\mu$ corresponds to the
ratio of the evaluation of $K$ and the sampling PDF. We then backpropagate gradients
through $\mu$, weighted by the loss gradient and $\verb|u|$ divided by the
\emph{detached} $\mu$. Here, the function \verb|detach(x)| disconnects its
argument from the automatic differentiation graph. This allows writing
Algorithm~\ref{alg:fredholm_grad} without duplicating the computation of $\mu$
once inside and once outside the \verb|ad_enabled()| block.

\section{Results}

\begin{figure}
    \centering
    \includegraphics[width=\linewidth]{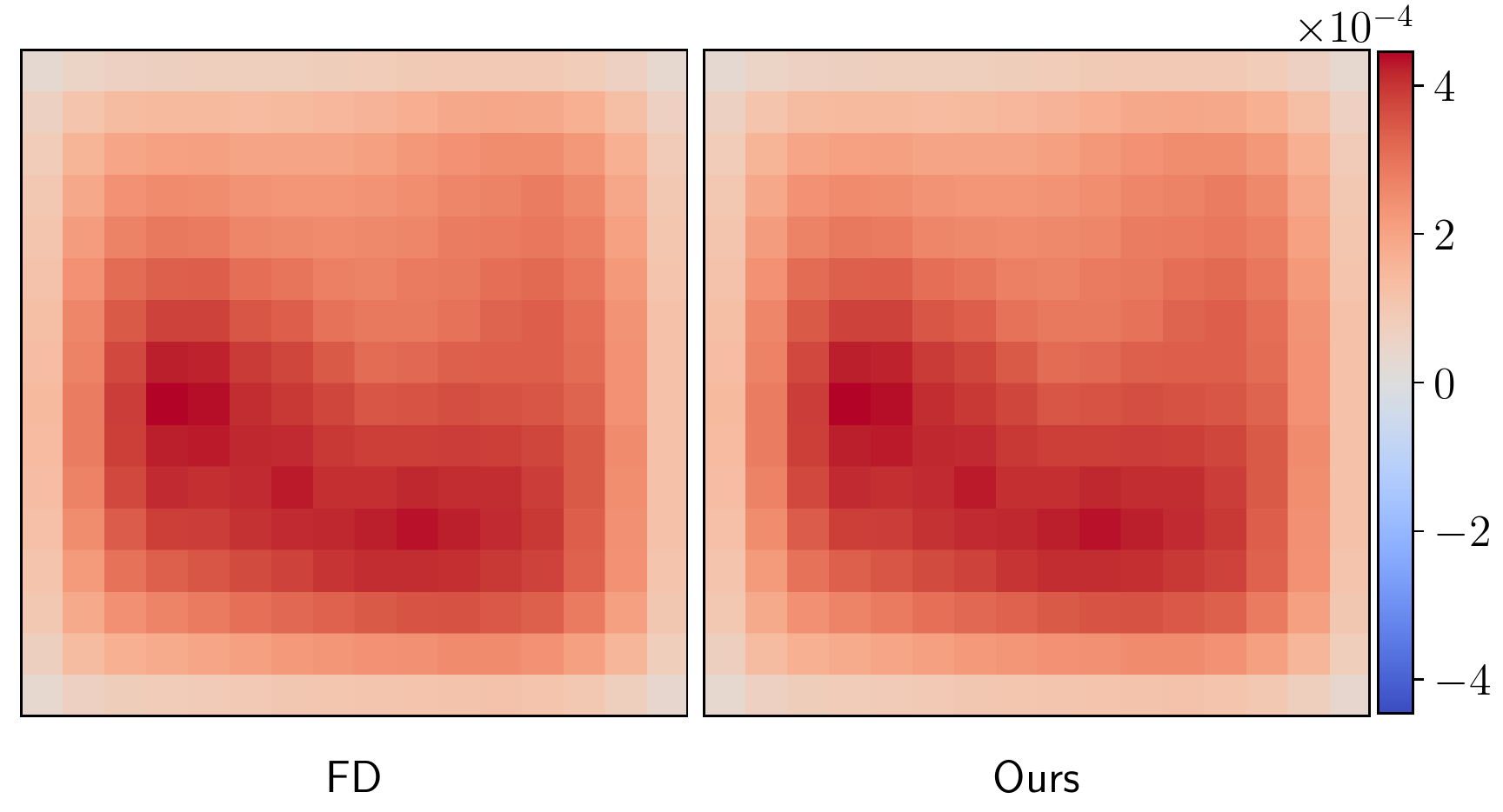}
    \vspace{-6mm}
    \caption{Validation of the source term gradient of the screened Poisson
            equation against finite differences (FD). The screening coefficient
            is set to 10. See Figure~\ref{fig:const_fd_input} for a
            visualization of the used source term and PDE solution.}
    \label{fig:const_fd_comparison}
\end{figure}

\begin{figure}
    \centering
    \includegraphics[width=\linewidth]{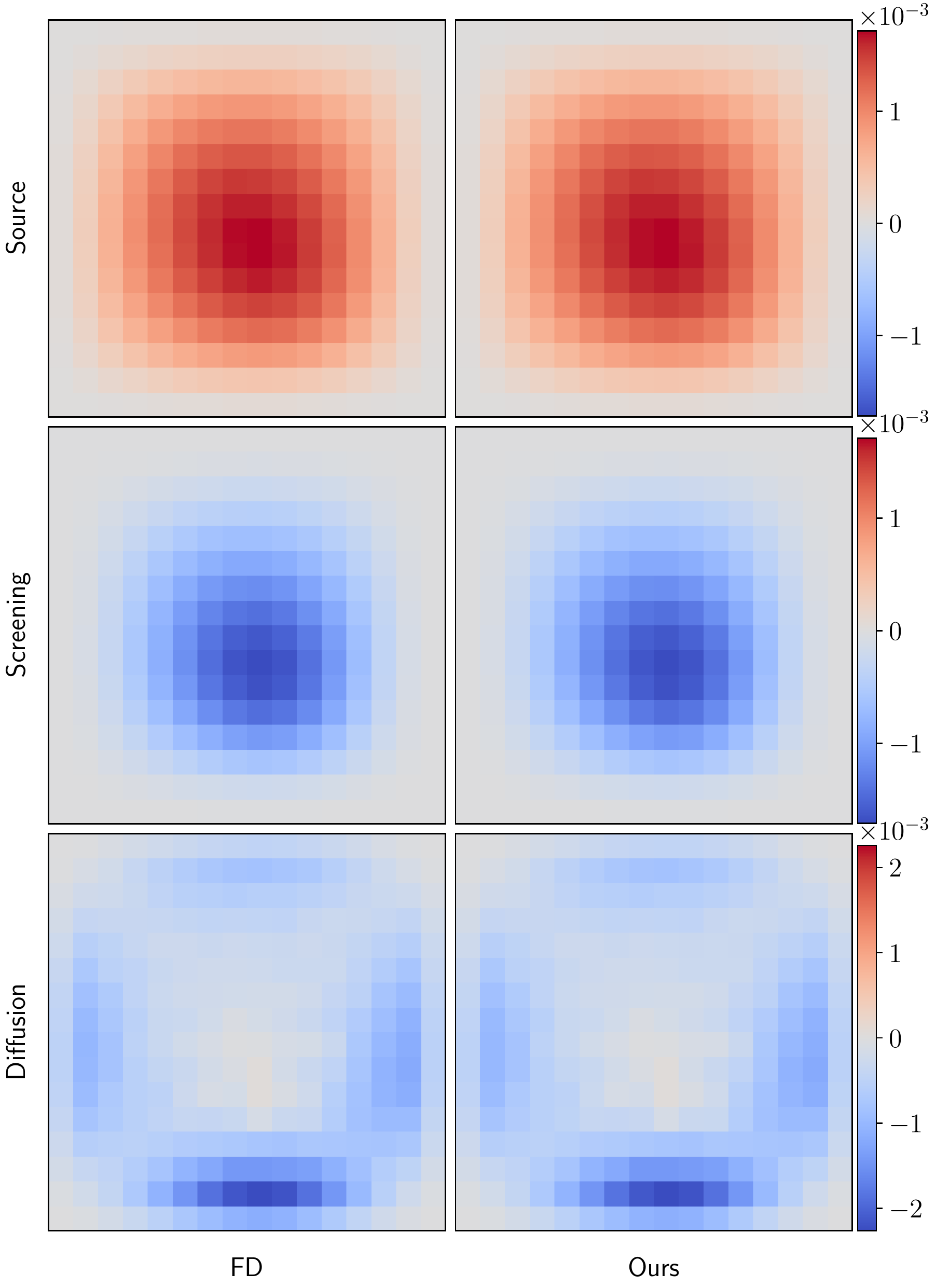}
    \vspace{-6mm}
    \caption{Comparison between our delta tracking gradient estimate and the
    finite differences reference for the various parameters of the general
    elliptic PDE.}
    \label{fig:fd_elliptic}
\end{figure}

\paragraph{Gradient validation}
We first validate the correctness of our gradient computations against finite
differences. In Figure~\ref{fig:const_fd_comparison}, we validate the source
term gradient for the screened Poisson solution. The objective function is
simply the $L_2$ norm of the solution image and the source function $f$ is a
cubically-interpolated $16 \times 16$ texture.

In Figure~\ref{fig:fd_elliptic}, we show a similar comparison for the general
elliptic PDE. The objective function is again the $L_2$ norm of the solution
image and we differentiate with respect to the source term $f$, screening
coefficient $\sigma$ and diffusion coefficient $\alpha$. The gradients
constructed using path replay perfectly match the finite difference reference,
which confirms that our estimator is indeed correct. The parameters for
this experiment are visualized in Figure~\ref{fig:elliptic_fd_input}.

\begin{figure*}[!h]
    \centering
    \includegraphics[width=\linewidth]{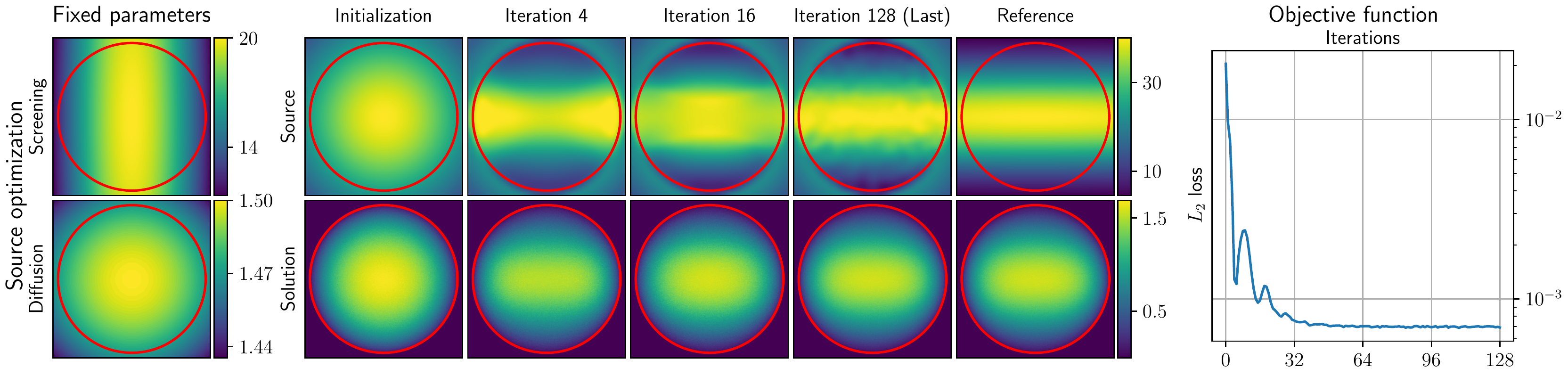}
    \includegraphics[width=\linewidth]{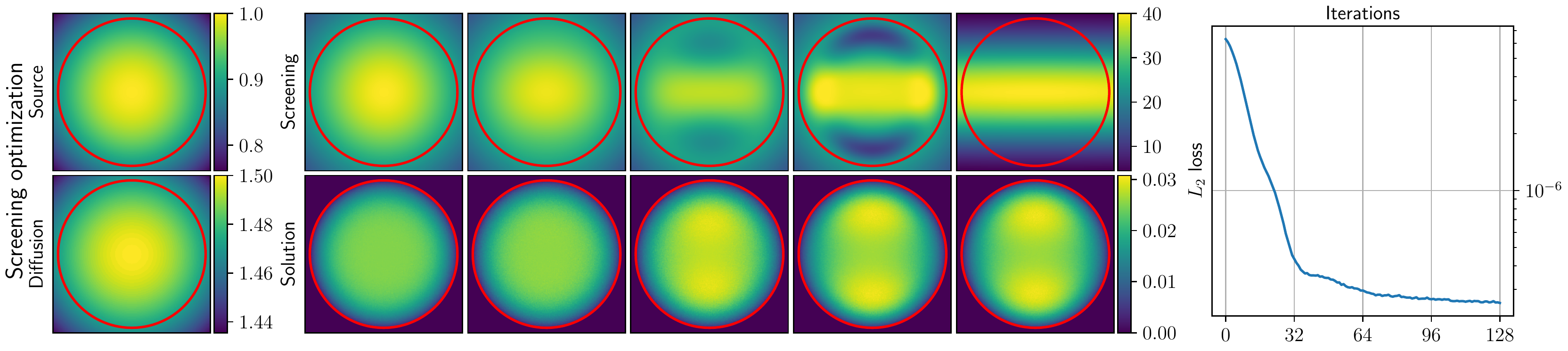}
    \includegraphics[width=\linewidth]{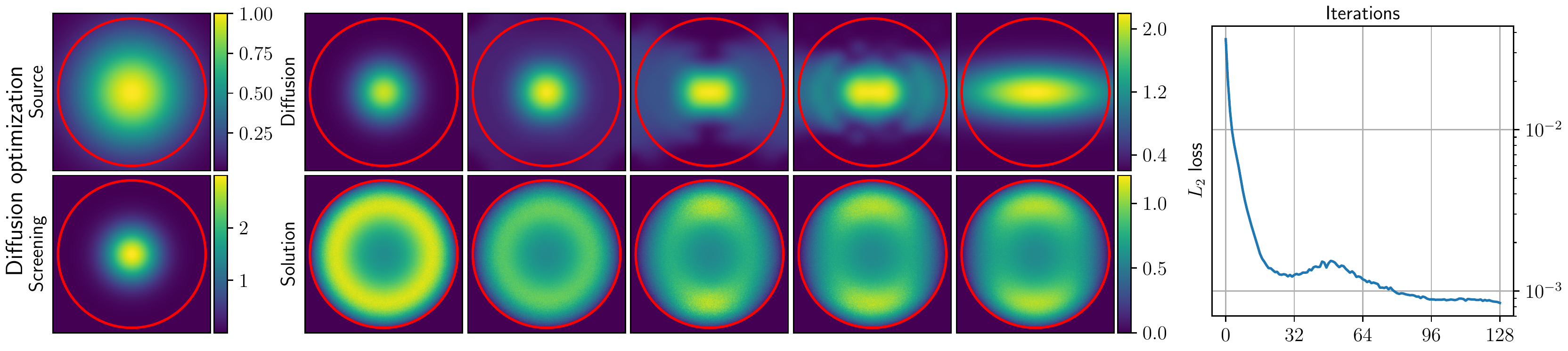}
    \vspace{-7mm}
    \caption{Optimization results for optimizing source term, screening
            coefficient and diffusion coefficient using the differential delta
            tracking WoS.}
    \label{fig:results}
\end{figure*}

\paragraph{Optimizations}
We demonstrate the use of our gradient estimators on a few example problems in
2D. We optimize various coefficients to minimize the $L_2$ error compared to the
reference. In each experiment we optimize one parameter represented by a regular
2D texture.

Figure~\ref{fig:results} shows the optimization results and plots of the
objective function values for three example problems. For all three problems,
the error decreases as the optimization progresses. However, this does not
necessarily guarantee that the recovered parameters match the reference
parameters. Although some inverse PDE problems admit unique
solutions~\cite{isakov_2006}, the elliptic problems discussed here are
inherently ambiguous, so any practical application most likely will require
using a prior or regularization.

The reconstruction of the source coefficient is the least vulnerable to these
ambiguities. As seen from the result in Figure~\ref{fig:results}, the optimized
source coefficient is almost identical to the reference source image in the
region of interest. This is consistent with observations made in differentiable
rendering, where emissive representations (e.g., neural radiance
fields~\cite{mildenhall2020nerf}) are easier to optimize than general
physically-based scene representations.

The optimization of the screening coefficient is partially successful in the
sense that the general shape of the reference screening $\sigma$ texture is
recovered. The reconstruction of the diffusion coefficient $\alpha$ appears to
be the least convex of our examples. In our experiments with different setups,
we observed that the diffusion coefficients can evolve in a way that the overall
shape of the texture is quite different from the reference coefficient, although
the optimized and the reference solutions are comparable. In all optimizations,
the solution of the PDE after optimization closely matches the reference.

\section{Conclusion and Future Work}
The main objective of this paper was to build differentiable Monte Carlo
estimators for solving inverse problems of PDEs. We proposed differential walk
on spheres algorithms for the types of PDEs discussed by
\citeauthor{Rohan22}~\shortcite{Rohan20, Rohan22}. Our methods are unbiased and
require a constant amount of memory. Their complexity is linear in the number of
sphere steps. We also provided some preliminary optimization results which
indicate that the differentiable Monte Carlo PDE solvers have the potential to
solve a range of inverse PDE problems.

As solving inverse problems of PDEs using differentiable Monte Carlo is a new
approach, there are many future directions to explore. It would be interesting
to also compute gradients with respect to the domain boundary. This could be
done using an attached version of PRB, as previously demonstrated for specular
surfaces~\cite{Vicini2021PathReplay}. Further development of inverse Monte Carlo
PDE methods would benefit from experiments on real-world problems, which could
highlight where our methods still lack. In particular, it would be interesting
to combine our approach with application-specific priors and regularization to
overcome local minima and ambiguities.

\begin{acks}
  We thank Rohan Sawhney and Dario Seyb for sharing the implementation of their
  primal solvers with us. We further thank Ziyi Zhang for proofreading, This
  research was supported by the Swiss National Science Foundation (SNSF) as part
  of grant 200021\_184629.
\end{acks}

\bibliographystyle{ACM-Reference-Format}
\bibliography{references}

\appendix

\section{Green's Functions and Poisson Kernels}
\label{appendix:green}
For completeness, we recapitulate the formulats for for the Green's functions
and the Poisson kernels in 2D and 3D. The Green's functions on
the ball $B(x)$ are:
\begin{align*}
    G_{2D}^{\sigma} (x,y) &= \frac{1}{2\pi} \left[ K_0(r\sqrt{\sigma}) -
    \frac{K_0(R\sqrt{\sigma})}{I_0(R\sqrt{\sigma})}I_0(r\sqrt{\sigma})\right] \\
    G_{3D}^{\sigma} (x,y) &= \frac{1}{4\pi} \left[  \frac{e^{-r\sqrt{\sigma}}}{r} -
    \frac{e^{-R\sqrt{\sigma}}}{R} \Bigg( \frac{\text{sinh}(r\sqrt{\sigma})}{r\sqrt{\sigma}}
    \frac{R\sqrt{\sigma}}{\text{sinh}(R\sqrt{\sigma})}  \Bigg)\right]
\end{align*}
where $r$ is the Euclidian distance between $x$ and $y$, R is the radius of the
ball $B(x)$, $I_0$ and $K_0$ are the zeroth order modified Bessel functions of
the first and the second  kind. The integral of the Green's function over the
ball $B(x)$ is used to compute the PDF of sampling proportional to it:
\begin{align*}
    |G_{2D}^{\sigma} (x)| = \int_{B(x)} G_{2D}^{\sigma} (x, y) \diff y &=
    \frac{1}{\sigma} \left[ 1 - \frac{1}{I_0(R\sqrt{\sigma})}\right] \\
    |G_{3D}^{\sigma} (x)| = \int_{B(x)} G_{3D}^{\sigma} (x, y) \diff y &=
    \frac{1}{\sigma} \left[ 1 - \frac{R\sqrt{\sigma}}{\operatorname{sinh}(R\sqrt{\sigma})}\right] \\
\end{align*}
The Poisson kernel is the derivative of the Green's function along the normal of
the boundary:
\begin{align*}
    P_{2D}^{\sigma} (x,z) &= \frac{1}{2\pi R} \left[ \frac{1}{I_0 (R \sqrt{\sigma})}\right] \\
    P_{3D}^{\sigma} (x,z) &= \frac{1}{4\pi R^2} \left[ \frac{R\sqrt{\sigma}}{\text{sinh}(R\sqrt{\sigma})}\right]
\end{align*}
The Poisson kernel is equal to $\frac{1 - \sigma |G^{\sigma}(x)|}{|\partial
B(x)|}$, which is a key property used in the derivation of the delta tracking
WoS.

\section{Input configurations of the gradient validation}
\label{appendix:input_conf}
\begin{figure}[!h]
    \centering
    \includegraphics[width=0.85\linewidth]{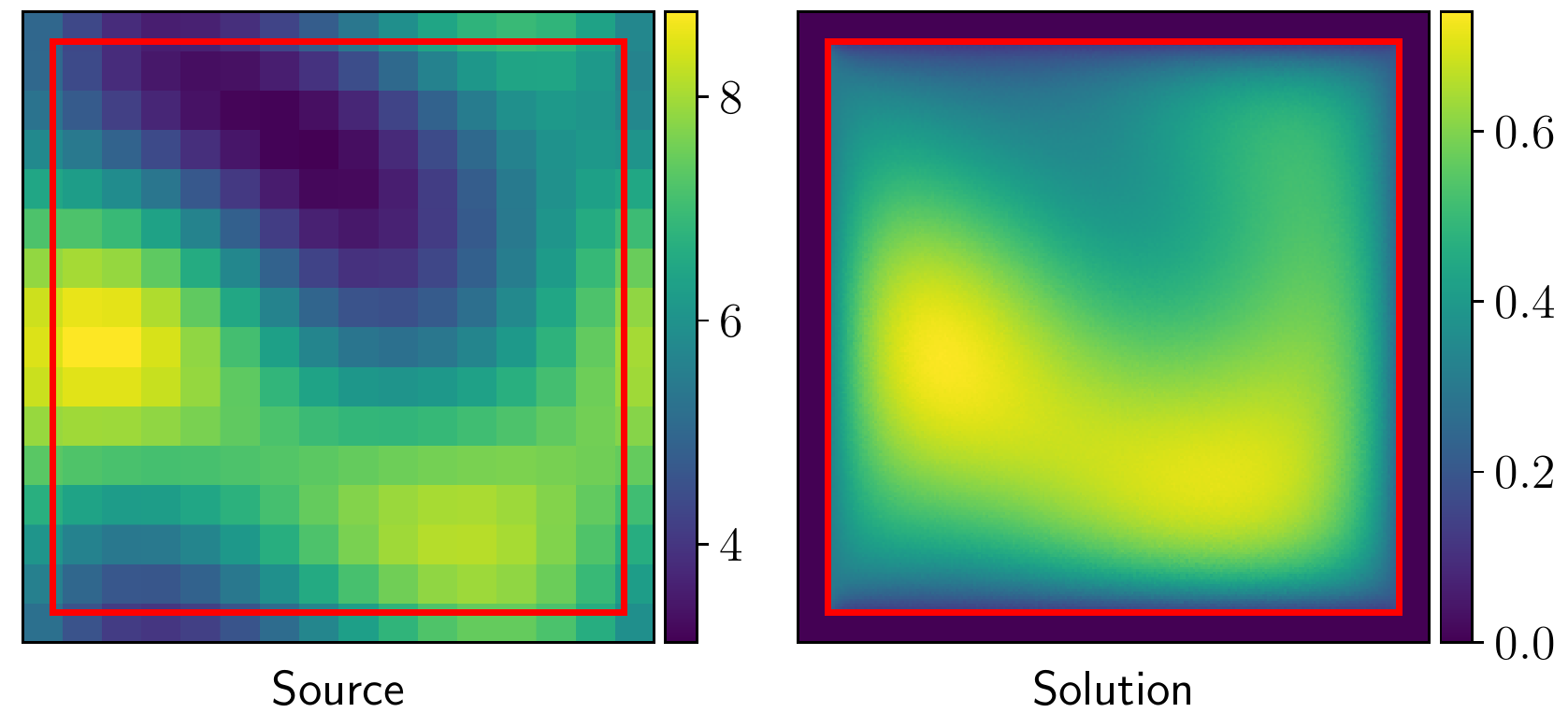}
    \vspace{-3mm}
    \caption{Input configuration of the finite difference test for the screened
             Poisson equation. Boundaries are indicated with red lines and all
             boundary values are set to zero. The screening coefficient $\sigma$
             is set to 10.}
    \label{fig:const_fd_input}
\end{figure}

\begin{figure}[!h]
    \centering
    \includegraphics[width=0.85\linewidth]{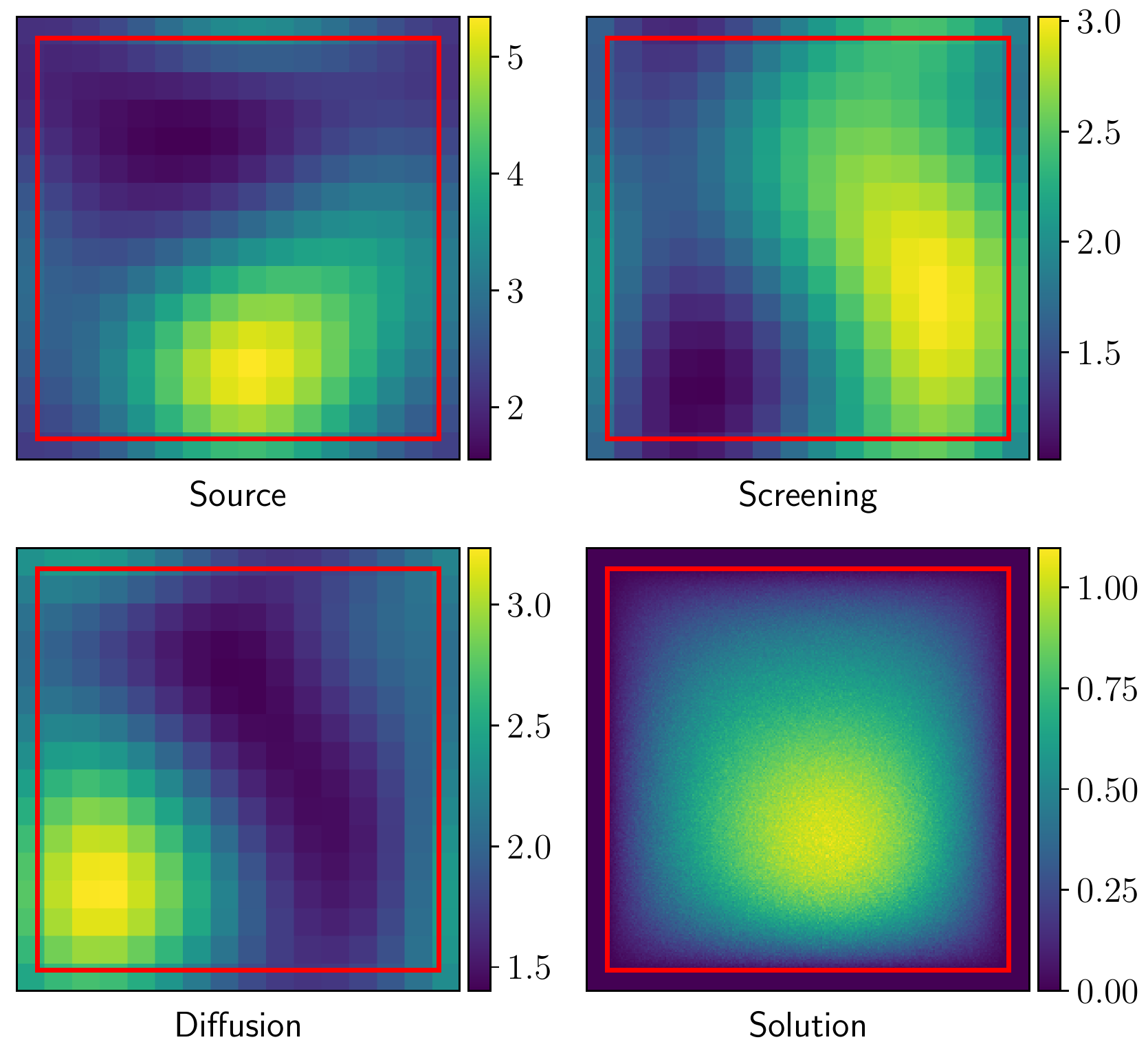}
    \vspace{-3mm}
    \caption{Input configuration of the finite difference test of the delta
             tracking WoS. Boundaries are indicated with red lines and all
             boundary values are set to zero.}
    \label{fig:elliptic_fd_input}
\end{figure}
\newpage
\onecolumn
\section{Parameter gradients}
\label{appendix:opt}

\newcommand\optsize{.29}

\begin{figure}[!h]
    \centering
    \includegraphics[width=\textwidth]{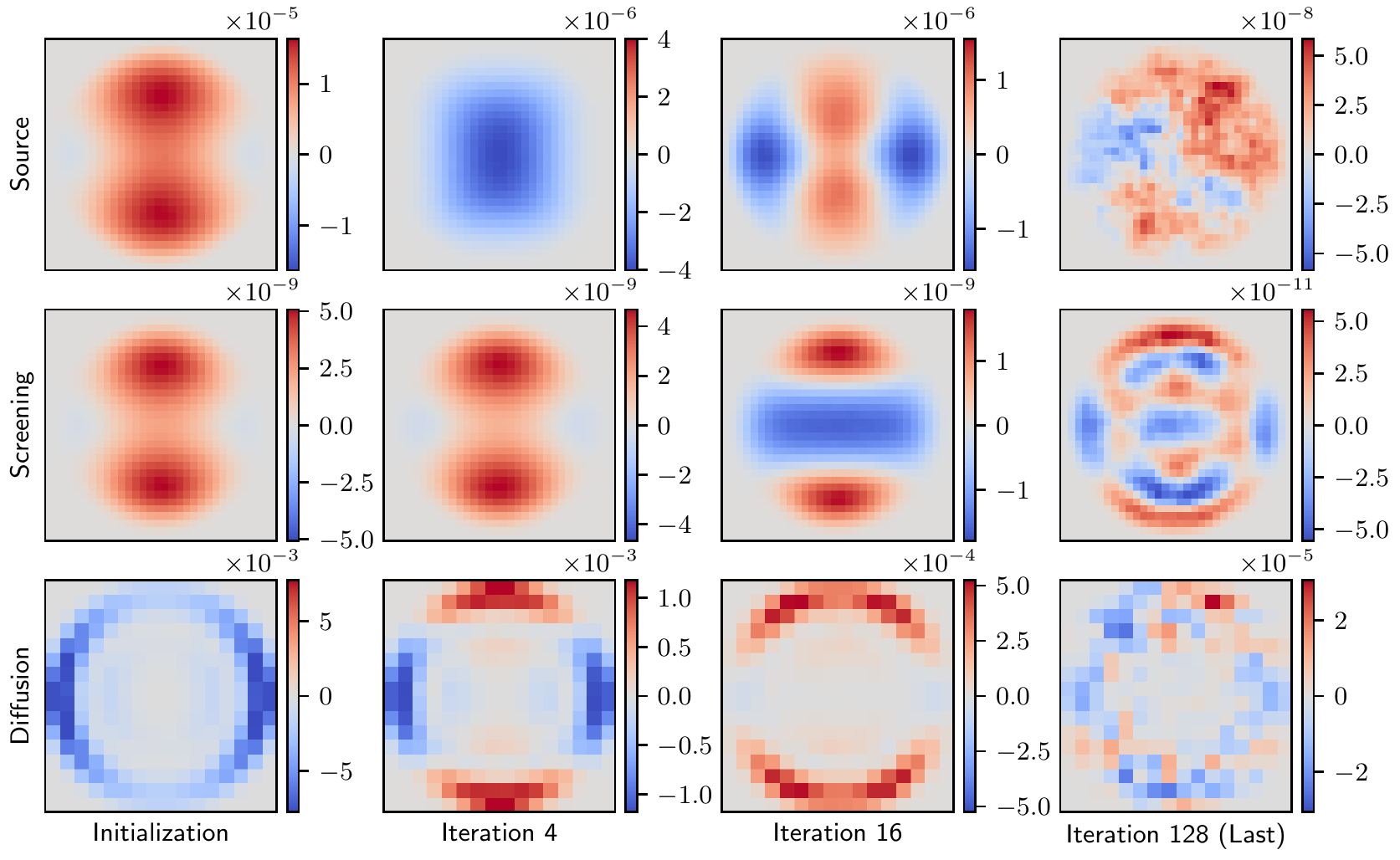}
    \vspace{-7mm}
    \caption{Visualization of the parameter gradients for different optimization
            iterations. As the optimizations progress, the objective function
            value and hence the magnitude of the parameter gradients decreases.}
    \label{fig:grad_it}
\end{figure}

\end{document}